# An Analysis of Conversational Volatility During Telecollaboration Sessions for Second Language Learning

**Aparajita Dey-Plissonneau**[1], **Hyowon Lee**[2], **Mingming Liu**[2],
**Vyoma Patel**[2], **Michael Scriney**[2], **Alan F. Smeaton**[2]
[1]School of Applied Languages and Intercultural Studies and
[2]Insight Centre for Data Analytics,
Dublin City University, Dublin 9, Ireland

*Abstract*

*Tandem telecollaboration is a pedagogy used in second language learning where mixed groups of students meet online in videoconferencing sessions to practice their conversational skills in their target language. We have built and deployed a system called L2 Learning to support post-session review and self-reflection on students' participation in such meetings. We automatically compute a metric called Conversational Volatility which quantifies the amount of interaction among participants, indicating how dynamic or flat the conversations were. Our analysis on more than 100 hours of video recordings involving 28 of our students indicates that conversations do not get more dynamic as meetings progress, that there is a wide variety of levels of interaction across students and student groups, and the speaking in French appears to have more animated conversations than speaking in English, though the reasons for that are not clear.*

# 1. Introduction

One of the most important aspects of second language learning, known as L2 learning, involves practicing conversations and dialogue, ideally with native speakers of the target language. This helps to reinforce the learner's confidence in the target language, expands vocabulary and exposes the learner to experiences of the cultural norms and practices of that target language.

In pre-COVID times, practicing conversational second language learning was difficult unless the learner relocated to a country where that language was spoken, which happened often through student exchanges funded by the Erasmus program, or through local meetups and gatherings. The concept of telecollaboration or virtual exchange whereby students would meet with native language speakers of their target language via videoconferencing calls had been used for a long time pre-COVID and its use was has long been a part of virtual exchange pedagogy (O'Rourke and Stickler, 2017). Tandem collaboration involves a reciprocal arrangement whereby the groups paired up for telecollaboration sessions spend half the call duration speaking in one language which would be the target language of one group and the native language of the other, and at the halfway point they would change language and the roles would be reversed.

Now, as a result of restrictions introduced by the pandemic, we have had more than 2 years of experience with Zoom, Microsoft Teams, Webex and other supports for teaching and learning, for social gatherings among friends and family, and for business purposes. As a result we are now more comfortable and familiar with the use of video conferencing in our day-to-day activities, and more open to its use in a variety of contexts.

In this paper we outline how we developed and deployed a software platform to support second language learning via telecollaboration in our University and in partnerships with 7 other Universities in Europe. This covered second language learning of English, French, German, Spanish and Italian. We automatically extracted a variety of metrics from each of the +400 telecollaboration meetings which were then presented to student participants to aid them in self-reflection on their participation in these online meetings.

One of the metrics we presented was conversational volatility, a measure reflecting the dynamic vs. static nature of conversational interactions. This paper presents the results of the conversational volatility metric for 114 telecollaboration meetings for a class of 28 students from our University learning French as a second language, partnered with students from Belgium learning English. Groups of up to 4 students were created and the groups met weekly for videoconferencing sessions for up to 6 weeks in a row. We explore whether there is variety in the level of conversational interactions across different groups as reflected in the conversational volatility metric and whether the dynamicity of the dialogue during the online sessions changed from one week to the next.

**2. Background**

Telecollaboration in language learning is not a new concept and a history of its development is presented in (Dooly & O'Dowd, 2018). More recently, a bibliometric analysis of 254 research articles in (Barbosa & Ferreira-Lopes, 2021) presented a summary of the field including an exploration into the benefits and drawbacks of different technologies that can support it.

In a recent editorial position paper, Colpaert (2020) took issue with the terminology used to describe this pedagogy and the shift to the use of the term "virtual exchange" but does expand on the many ways in which "telecollaboration affords many more activities than its physical counterparts", thus further strengthening arguments in favour of telecollaboration. What we are interested in here is how technology can be, and has been used, to support the specific needs of telecollaboration.

The bibliometric study presented in (Barbosa & Ferreira-Lopes, 2021) addressed this by looking at the most-used technologies and found that videoconferencing systems like Skype, virtual worlds like Second Life, social media platforms like Facebook, storytelling wikis and blogs, were the main tools used, drawing much of its evidence from (Avgousti, 2018). There are also quite a few software platforms specifically built to offer support for tandem language learning including HelloTalk, Babbel.com, SpeakPlus and others.

However what is common to the specialist platforms, the videoconferencing systems and the other tools used is that they do not support post-session reviews. In particular, there is nothing to specifically support students' self-reflection on their telecollaboration tandem meetings, which is what we address in this paper.

**3. The L2 Learning System**

The L2 Learning system (Dey-Plissonneau et al, 2021a) is built on top of Zoom and uses the Zoom audio transcripts which are an encoding of Zoom's timed speech recognition. Students hold their Zoom meetings typically in groups of 3 or 4 with half the meeting being in English and half in the second language. After each meeting with students when Zoom has completed its processing and speech transcription, students share the link to their video recording and upload the transcript file (in VTT format) to the L2 Learning system. From this we generate a visualisation of the telecollaboration meeting as shown in Figure 1.

The visualisation shown in Figure 1 which is not one from the student telecollaborations, shows a meeting between 3 participants. The features include a hotlinked timeline as a series of vertical bars in blue, red and yellow showing who spoke when and for how long as well as the overall % participation for each participant. A chord graph on the right of the screen indicates the cumulative sequence of who followed who in the conversation. A metric we

*Conversational Volatility During Tandem Telecollaboration*

call conversation volatility, described in the next section, is also shown for the overall and for each half of the meeting as a bar chart on the bottom left. The conversation flow chord graph is useful to illustrate when a subset of participants dominate the Zoom call by having their own conversations among themselves and 1 or more of the remainder are left out of the dialogue. This can happen for any of several reasons, including when a participant is not comfortable speaking the language of the conversation at that point in the Zoom call. The colour coded utterances on the timeline are hotlinked so clicking on any of them starts video playback shown in the middle of the screen with the headshots as a gallery view of the 3 participants in this case (with faces blurred).

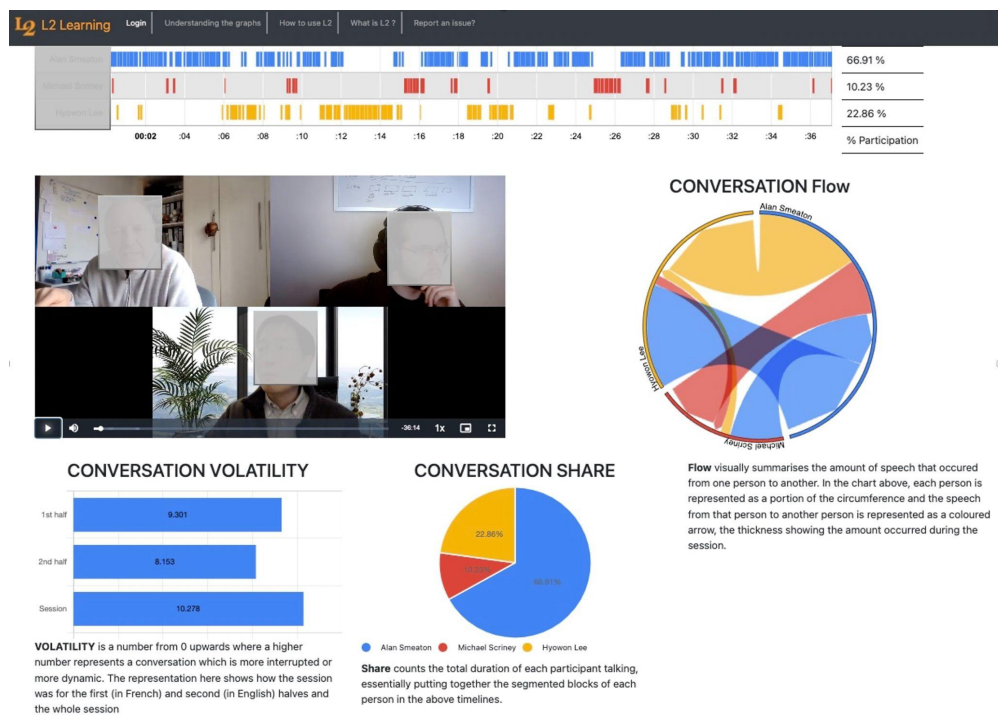

Figure 1: Screengrab from L2 Learning System

We have been using the L2 Learning system for three full semesters in collaborations between 2 English-speaking Universities and Universities speaking French, Spanish, German and Italian (Dey-Plissonneau et al, 2021b). For this paper we focus on data gathered from 28 students in an English-speaking country taking a 1-semester French language course at intermediate level in Autumn 2021. These students were grouped with 36 students in a French-speaking University in Belgium who were learning English and a total of 114 Zoom tandem telecollaboration meetings of average duration 58 minutes were recorded and analysed.

## 4. Conversational Volatility

When analysing a telecollaboration meeting in an automatic and scalable way, features like total speaking time are useful but what we would really like to identify is the amount of interaction or turn-taking during the Zoom meeting by the meeting as a whole as well as by each participant. To address this we introduced conversation volatility as a metric. As shown in (Guydish and Fox Tree, 2021) while the characteristics of good conversations are complex, and even more so with new online communication technologies, the rationale is that the more that participants engage in a conversation, the better the experience will be for all. Thus an online conversation with lots of interaction and interruption from multiple participants will have a higher volatility measure and result in a more enjoyable experience for all participants than a conversation which is a flat series of monologues with low conversational volatility.

Historical volatility (Hong and Lee, 2017), (Somarajan et al., 2019) is a statistical measure which is widely used in applications in economics and finance. It is used by analysts and stock traders as part of the creation of financial investing strategies. Historical volatility is formally defined as the degree of variation of values of some continuous time series over time, usually measured by the standard deviation of daily changes in stock prices.

If we apply the historical volatility metric to turn taking for a telecollaboration meeting, that will indicate whether the dialogue was truly interactive and composed of shorter and longer utterances mixed such as when people interrupt each other, or whether it consisted of long monologues with likely tedious turn-taking. Conversational volatility can attribute scores to the dialogue as a whole or to individual participants. Here we compute conversational volatility for the first and the second halves independently as well as for the whole meeting. This would indicate whether there was more interaction in the French or English speaking parts of the meetings.

## 5. Results

A total of 114 telecollaboration Zoom meetings involving the 28 students from our University form the data for analysis of conversational volatility in this paper. The average Zoom meeting length was just under 58 minutes with little variation either side of that, so students were consistent in keeping to the one hour recommended meeting duration. Students were asked to spend the first half of the meeting in French and at about the mid-way point to switch to English. When we manually annotated the turnover point for the 114 meetings we found that on average this point was within 3 minutes 44 seconds of the actual midway point of the recorded meeting. There were 2 of 114 meetings that spent much longer in one language before changing and when we remove those the changeover point dropped to within 3 minutes of the actual midpoint. This means that when calculating conversational volatility for the French and English parts of telecollaboration meetings we can use the half-way point to determine the point of language changeover without much error. Our annotation found that 93 of the 106 meetings started in French followed by English and 13 had the opposite so we would need some form of language identification if we are to completely automate the calculation of conversational volatility, though for the analysis presented here we use the manual annotation of which half was in French and which was in English.

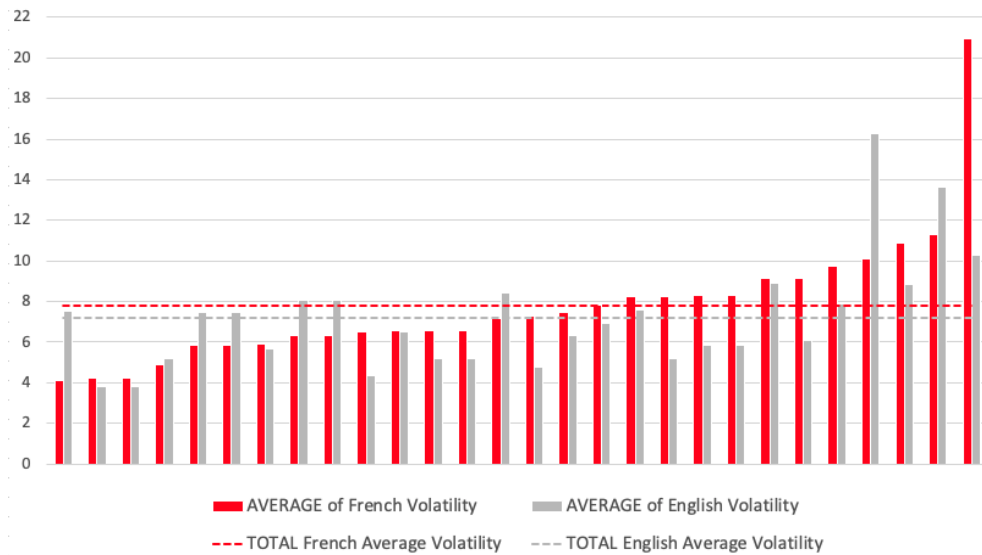

*Figure 2: Average conversation volatility per group speaking each language (groups ordered by increasing volatility for French sessions)*

Figure 2 shows the average conversation volatility measures for each of 28 students for the English-speaking and French-speaking halves of their meetings. There is a small spread of average values though with some extreme values, both very dynamic and very static. Of the 28 students in groups, 19 had higher volatility for French than for English parts and 9 had higher volatility for English and the average for French (7.8) was greater than for English (7.2). Given that it is the same sets of students in a group speaking English and French, does this difference in languages indicate that speaking in French is more dynamic than in English? We do not yet have enough evidence for this and it is a topic for further investigation.

We then looked at how the volatility measure changes for 19 of the 28 students as they progress from one meeting to the next for the French-speaking parts of their meetings and this is shown in Figure 3. We chose these 19 students as they had missed fewer than their peers so had greater contiguity. The x-axis labels indicate the numbers of students who had 1 meeting (19 students), 2 meetings (19 students) and so on up to 6 meetings (3 students). The dotted red line shows the average volatility measure for first, second, etc. meetings.

Figure 3 reinforces what we saw in Figure 2, that there is a spread of values for almost all students with some extremes. Almost all students they have ups and downs throughout their meeting progressions, with little flatlining on the graph. We also note that there is no overall increase in conversational volatility as meetings progress from first to last and this is shown in the averaged value per meeting (dotted red line) as well as in entries for individual students.

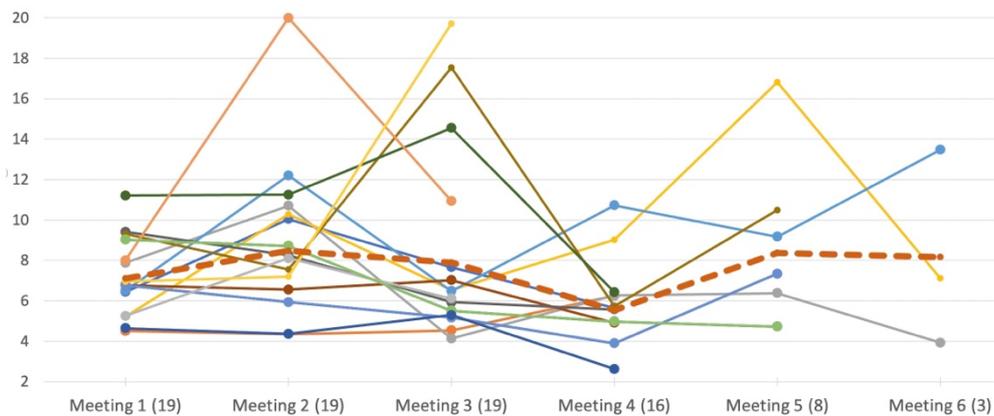

*Figure 3: Conversational volatility per group for French-speaking parts of online meetings*

Finally, we calculated conversational volatility measures for individual student participation in meetings. The values for volatility for individuals are less than those for whole groups because of smaller numbers of utterances. For 87 meetings involving these 19 students with 68 hours of video recordings and transcriptions, almost half had higher volatility for their French speaking contributions and half had higher volatility for their native English speaking contributions but the French parts did have higher volatility levels on average.

## 6. Conclusions

In this paper we report on the levels of conversational interaction which took place as part of tandem telecollaboration for 28 students learning French as a second language, partnering with French-speaking students learning English and using the L2 Learning system. We introduced a measure of conversational dynamics called conversational volatility which quantifies the amount of interaction among participants at group and individual levels.

Our analysis shows that levels of conversational dynamics does not increase as students progress through their weekly tandem telecollaboration sessions and that video conversations in French seem to have more interaction and turn taking than in English. We found variety in the levels of interaction for different students with some more animated and interactive than others. Finally we observed variety in the levels of interaction for students across their own meetings. The reasons for these observations are all topics for future investigation.

**Acknowledgements**
This publication has emanated from research supported by Science Foundation Ireland (SFI) under Grant Number SFI/12/RC/2289_P2 (Insight SFI Research Centre for Data Analytics), co-funded by the European Regional Development Fund.

**References**

Avgousti, M.I. (2018) Intercultural communicative competence and online exchanges: a systematic review, *Computer Assisted Language Learning*, 31:8, 819-853, doi: 10.1080/09588221.2018.1455713

Barbosa M.W. & Ferreira-Lopes, L. (2021) Emerging trends in telecollaboration and virtual exchange: a bibliometric study, *Educational Review*, doi: 10.1080/00131911.2021.1907314

Colpaert, J. (2020) Editorial position paper: how virtual is your research?, *Computer Assisted Language Learning*, 33:7, 653-664, doi: 10.1080/09588221.2020.1824059

Dey-Plissonneau, A., Lee, H., Pradier, V., Scriney, M. & Smeaton, A.F. (2021a). The L2L system for second language learning using visualised Zoom calls among students. *European Conference on Technology Enhanced Learning* (pp. 409-413). Springer, Cham. doi: 10.1007/978-3-030-86436-1_45

Dey-Plissonneau, A., Lee, H., Scriney, M., Smeaton, A.F., Pradier, V. & Riaz, H. (2021b). Facilitating reflection in teletandem through automatically generated conversation metrics and playback video. *CALL and professionalisation: short papers from EUROCALL 2021*, 88. doi: 10.14705/rpnet.2021.54.1314

Dooly, M., & O'Dowd, R. (2018). Telecollaboration in the foreign language classroom: A review of its origins and its application to language teaching practice. *M. Dooly and R. O'Dowd (Eds.)*, 11-34

Guydish, A.J, Fox Tree, J.E. (2021). Good conversations: Grounding, convergence, and richness. *New Ideas in Psychology*, 63, doi: 10.1016/j.newideapsych.2021.100877.

Hong, Y., & Lee, Y. J. (2017). A general approach to testing volatility models in time series. *Journal of Management Science and Engineering*, 2(1), 1-33. doi: 10.3724/SP.J.1383.201001

O'Dowd, R. (2021) Virtual exchange: moving forward into the next decade, *Computer Assisted Language Learning*, 34:3, 209-224, doi: 10.1080/09588221.2021.1902201

O'Rourke, B. and Stickler, U. (2017). "Synchronous communication technologies for language learning: Promise and challenges in research and pedagogy". *Language Learning in Higher Education*, 7(1), 1-20. doi: 10.1515/cercles-2017-0009

Somarajan, S., Shankar, M., Sharma, T., & Jeyanthi, R. (2019). Modelling and analysis of volatility in time series data. In *Soft Computing and Signal Processing* (pp. 609-618). Springer, Singapore. doi: 10.1007/978-981-13-3393-4_62